\documentclass[a4paper,11pt]{article}
\pdfoutput=1

\usepackage{jinstpub} 
\usepackage{slashbox} 
\newcommand{\diff}{\mathop{}\!\mathrm{d}} 
\newcommand*\V[1]{\mathbf{#1}} 
\usepackage{graphicx}
\graphicspath{{fig/}}
\usepackage{siunitx}

\title{Discrimination of nuclear and electronic recoil events \\ using plasma effect in germanium detectors}

\author[a] {W.-Z. Wei,}
\author[a] {J. Liu}
\author[a,b,1]{and D.-M. Mei\note{Corresponding author.}}

\affiliation[a]{Department of Physics, The University of South Dakota,\\414 E. Clark Street, Vermillion, South Dakota 57069, USA}
\affiliation[b]{School of Physics and Optoelectronic, Yangtze University, \\ 1 Nanhuan Street, Jingzhou 434023, China}

\emailAdd{Dongming.Mei@usd.edu}

\abstract{We report a new method of using the plasma time difference, which results from the plasma effect, between the nuclear and electronic recoil events in high-purity germanium detectors to distinguish these two types of events in the search for rare physics processes. The physics mechanism of the plasma effect is discussed in detail. A numerical model is developed to calculate the plasma time for nuclear and electronic recoils at various energies in germanium detectors. It can be shown that under certain conditions the plasma time difference is large enough to be observable. The experimental aspects in realizing such a discrimination in germanium detectors is discussed.}


\arxivnumber{1605.05244} 

\begin{document}
\maketitle
\flushbottom

\section{Introduction}
\label{s:intr}
In the detection of dark matter or neutrino-nucleus coherent scattering induced nuclear recoil events (NRs) with high-purity germanium detectors such as SuperCDMS~\cite{cdms}, CoGeNT~\cite{cog} and CEvNS~\cite{schol} experiments, the main background comes from the electronic recoil events (ERs) produced by natural radioactivity. The capability of
discriminating NRs from ERs is crucial in reducing the background to reach a better sensitivity for those experiments. The germanium cryogenic bolometers such as CDMS~\cite{cdms}- and EDELWEISS~\cite{bron}-type detectors provide excellent discrimination power between NRs and ERs by measuring ionization yield which is the ratio of the ionization energy to the phonon energy. However, the bolometers must be operated in a temperature range of milli-Kelvin, which demands high cost for large detectors that are needed for the next generation ton-scale experiments. Compare with a cryogenic bolometer, a germanium detector operated at liquid nitrogen temperature (77 Kelvin) is relatively simple and does not require complex cooling systems. Thus, it would be quite attractive if a generic germanium detector is capable of identifying NRs from ERs. Digital pulse shape analysis is an encouraging approach to the discrimination of ions with different mass numbers due to their difference in length and density of their ionizing tracks~\cite{caj, aaq, wde, tki, jba, ssk, gpa}. Similar difference is expected between NRs and ERs. This is because a nucleus is much heavier than an electron and the heavier particle generates ionization more densely along its path, we expect that electron-hole (e-h) pairs creation and  charge collection are different between NRs and ERs due to the differences of length and density of the ionization track. The rise time of the pulse shape is essentially governed by two effects:
\begin{enumerate}
\item{The drift time of the charge carriers that move along the electric field lines towards the corresponding electrode.} This drift time is called the charge transit time, which depends on the drift paths and the drift velocities for electrons and holes.

\item{The density of e-h pairs along the track of the particle.} A high density of charge carriers along the ionization track forms a plasma-like cloud of charges that shields the interior from the influence of the electric field.  Only those charge carriers at the outer edge of the cloud are subject to the influence of the electric field, and they begin to migrate immediately. This plasma-like cloud expands radially due to diffusion of charge carriers and is gradually eroded away until the charges at the interior are finally subject to the applied field and also begin to drift. The time needed for total disintegration of this plasma region is called the plasma time, which is known as the second component of the pulse rise time. The plasma time depends on the initial density and radius of the plasma-like cloud, on the diffusion constant for charge carriers, and on the strength of electric field~\cite{jba,wse, ecf,ika}.  Both drift time and plasma time are responsible for the pulse rise time in the charge collection. The question is to determine the difference in the ionization length and density of e-h pairs between NR and ER events.
\end{enumerate}

In the dual-phase xenon detectors for the direct detection of dark matter~\cite{LUX, xenon100, pandax}, the plasma time plays an important role in the recombination of electron-ion pairs~\cite{wangmei} as a function of recoil energy, type, and electric field. The anti-correlation between the charge and light yield clearly demonstrates the recombination probability as a function of the plasma time~\cite{wangmei}.

A direct observation of the plasma time in a silicon detector~\cite{rai} found a value of $\sim$3-5 ns for an alpha particle with energy about 5 MeV. In the calculation of the plasma time $\tau_{pl}$, England et al.~\cite{jba} have proposed a model which takes into account the dependence of the plasma time on the applied electric field $F$ and the density of the ionization track $\diff E/\diff x$:
\begin{equation}
\label{eq:plasma}
\tau_{pl} = \frac{\beta}{F} \sqrt{\frac{\diff E}{\diff x}},
\end{equation}
where $\beta$ stands for a normalization factor which is determined by the experiment.  This equation is also expressed by the following form which brings out the $Z-$ and $M-$dependence of projectile on $\tau_{pl}$:
\begin{equation}
\label{eq:plasma1}
\tau_{pl}(ns)  = n (MZ^{2})^{1/2}[B(\overline{\frac{1}{E}ln\frac{4m_{0}E}{MI}})]^{1/2}/F,
\end{equation}
where $B$ = 2$\pi e^{4}N_{0}Z/m_{0}A$ is the Bethe-Bloch constant, $M$ is the mass of the incident ion, $Z$ is the atomic number of the incident ion, $E$ is the energy for the recoiling ion, $I$ is the average ionization energy for the absorbing material, m$_{0}$ is the rest mass of electron and $n$ is a normalization constant that is determined experimentally. Seibt et al.~\cite{wse} used first principles assuming a diffusion process combined with a radial space-charge expansion of the plasma as electrons are removed  to calculate the plasma time. The result for silicon is:
\begin{equation}
\label{eq:plasma2}
t_{pl}(s) = 1.32 \times 10^{-10} (n_1E)^{1/3}/F,
\end{equation}
where $E$ is the energy of the recoiling ion, $n_{1}$ is the total initial number of charge carriers per unit length of track and $F$ is the applied reverse bias field.   

In the case of germanium detectors, the plasma effect has been studied experimentally using a coaxial detector~\cite{Baudis98}, which indicates a negligible effect. The negative result from the experiment using a coaxial detector can be simply understood as the small plasma effect was  washed out by much longer drift time. The theoretical consideration needs to be developed. It is believed that the formation of the plasma effect on the ionization track requires the ratio of the Debye screening radius, $\lambda_{D}$, to the radius of the ionization column, $r$, $\frac{\lambda_{D}}{r}$ = $\sqrt{\epsilon kT/4e^2\eta}$ $\leq$ 1~\cite{dolg}, where $\epsilon$ = 16 is the dielectric constant of germanium, $k$ is the Boltzmann constant, $T$ is the temperature, $e$ is electron charge, and $\eta$ is the number of ions per unit length of the ionization track. To evaluate the plasma effect induced by NRs in germanium, one can look into the the number density of charge carriers created by an incoming neutron for a given recoil energy.

When a neutron elastically scatters off a germanium nucleus and transfers a portion of its  kinetic energy to the germanium nucleus, the germanium nucleus is knocked off its lattice site and then loses its energy by colliding with electrons and nuclei within the detector. Therefore, this NR process involves a competition between energy transfer to atomic electrons and energy transfer to translational motion of an atom. The total rate  at which it loses energy with respect to distance ($\diff E/\diff x$) is dependent on the medium through which it travels, and it is also called stopping power. At low energies, the total stopping power of the germanium is divided between the electronic and nuclear stopping power. Electronic stopping power is the amount of energy per unit distance that the recoil nucleus loses to electronic excitation and ionization of the surrounding germanium atoms.  Nuclear stopping power is the energy loss per unit length that the recoil nucleus loses to atomic collisions which add to the kinetic energy of the germanium atoms, but do not result in internal excitation of atoms. The energy once given to electrons can be transferred back to atomic motion in a very slow process.  The ratio of electronic to nuclear stopping power depends on the recoil energy of the nucleus. If the recoil energy is very large, the portion of the nuclear stopping power would be smaller compared to the portion of the electronic stopping power. However, in the energy range of the recoil germanium atoms from neutron collisions, the nuclear stopping power plays a significant role in the energy loss of the recoil nucleus. J. Lindhard et al.~\cite{lind10} discussed the theory of energy loss for low energy nuclei in detail. For instance, the number density of ions per track length created by 1 keV NR event is about 2.8$\times$10$^{8}$/cm in germanium, which gives $\frac{\lambda_{D}}{r}<$0.1. Therefore, the plasma effect in germanium can be formed. 

Due to the difference in the stopping power between NRs and ERs, the plasma effects created by the ionization density are expected to be different between NR and ER events for a given recoil energy. The calculation of the plasma time must take into account a dynamic process in which the density of charge carriers, the ambipolar diffusion, the external electric field, and the charge drifting are all involved.  Thus, it is natural to consider numerical calculations with all physics parameters that are involved in the creation and erosion of the plasma effect.  

In this paper, the numerical calculation including the general equations, simplifications in calculation, study of mobility, evolution of various distributions and estimation of plasma time is presented in section~\ref{s:numc}, followed by the results of the numerical calculation in section~\ref{s:res}. The experimental consideration on measuring the plasma effect in germanium detectors is presented in section~\ref{s:expe}. Finally, the conclusions are summarized in section~\ref{s:conc}. 

\section{Numerical calculation}
\label{s:numc}
\subsection{General equations}
The current densities $\V{j}_{e/h}$ generated by the drift of electrons and holes, respectively, can be calculated as:
\begin{equation}
  \V{j}_e(\V{x},t) = -qn(\V{x},t)\V{v}_e(\V{x},t),
  \label{e:je}
\end{equation}
\begin{equation}
  \V{j}_h(\V{x},t) = qp(\V{x},t)\V{v}_h(\V{x},t),
  \label{e:jh}
\end{equation}
where $q$ is the elementary charge, $n$ and $p$ are the number densities of electrons and holes, respectively, $\V{v}_{e/h}$ are the saturated drift velocities of electrons and holes, respectively, and $(\V{x},t)$ denotes the location and time dependence of $\V{j}, n, p$, and $\V{v}$. The drift velocities can be calculated as:
\begin{equation}
  \V{v}_{e/h}(\V{x},t) = \mu_{e/h}\V{E}(\V{x},t),
  \label{e:driV}
\end{equation}
where $\mu_{e/h}$ are the drift mobilities of electrons and holes, respectively, and $\V{E}$ is the sum of both external and induced electric fields:
\begin{equation}
  \V{E} = \V{E}_\text{ex} + \V{E}_\text{in}.
  \label{e:netE}
\end{equation}
$\V{E}_\text{in}$ appears when electron and hole clouds do not overlap with each other completely, and can be calculated with \textit{Gauss's Law}:
\begin{equation}
  \nabla \cdot \V{E}_\text{in}(\V{x},t) = \frac{qp(\V{x},t)-qn(\V{x},t)}{\varepsilon_0\varepsilon_{\text{Ge}}},
  \label{e:gl}
\end{equation}
where, $\varepsilon_\text{Ge}=16$ is the relative permittivity for germanium and $\varepsilon_0$ is the free-space permittivity.

The differential continuity equation provides the relationship between the time evolution of charge carrier number densities ($n$ and $p$) and the current densities $\V{j}_{e/h}$:
\begin{equation}
  -q\frac{\partial n(\V{x},t)}{\partial t} = - \nabla \cdot \V{j}_e,
  \label{e:ce}
\end{equation}
\begin{equation}
  q\frac{\partial p(\V{x},t)}{\partial t} = - \nabla \cdot \V{j}_h.
  \label{e:ch}
\end{equation}
The diffusion of charge carrier clouds adds another term to eq.~\ref{e:ce} and \ref{e:ch}:
\begin{equation}
  -q\frac{\partial n(\V{x},t)}{\partial t} = - \nabla \cdot \V{j}_e + \nabla \cdot [D(n(\V{x},t))\nabla n(\V{x},t)],
  \label{e:cde}
\end{equation}
\begin{equation}
  q\frac{\partial p(\V{x},t)}{\partial t} = - \nabla \cdot \V{j}_h + \nabla \cdot [D(p(\V{x},t))\nabla p(\V{x},t)],
  \label{e:cdh}
\end{equation}
where $D$ is the collective diffusion coefficient for density $n$ or $p$ at location $\V{x}$.

Given initial number density distributions, $n_0$ and $p_0$, the current densities $\V{j}_{e/h}$ can be calculated with eqs.~\ref{e:je}--\ref{e:gl}. The number densities $n_1$ and $p_1$ after a small time interval $\diff t$ can be then calculated as:
\begin{equation}
  n_1 = -f(n_0)/q\diff t,
\end{equation}
\begin{equation}
  p_1 = f(p_0)/q\diff t,
\end{equation}
where $f()$ represent the right hand sides of eq.~\ref{e:cde} and \ref{e:cdh}. Such an operation can be repeated $N$ times until the distributions of $n_N$ and $p_N$ are clearly separated from each other in space:
\begin{equation}
  n_{i+1} = -f(n_i)/q\diff t, \ \ i=0, 1, 2,...N,
  \label{e:ie}
\end{equation}
\begin{equation}
  p_{i+1} = f(p_i)/q\diff t, \ \ \ i=0, 1, 2,...N.
  \label{e:ih}
\end{equation}
The plasma time $t_\text{pl}$ can be then estimated as:
\begin{equation}
  t_\text{pl} = N\diff t.
  \label{e:tp}
\end{equation}

\subsection{Simplification}
\begin{figure} [htbp]
  \centering
  \includegraphics[clip,width=0.6\linewidth]{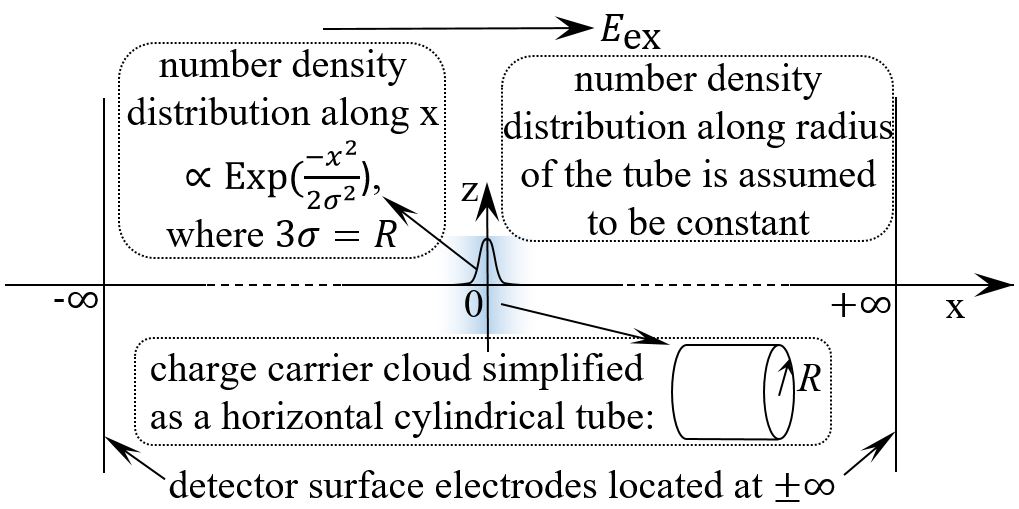}
  \caption{Initial setup for the numerical calculation in a planar germanium detector.}
  \label{f:cfg}
\end{figure}

In the case of a high-purity planar germanium detector with a constant high voltage applied to its left and right surface electrodes, as shown in figure~\ref{f:cfg}, the three-dimensional vector equations can be reduced to one dimensional ones with the following simplifications: (1) since the original size of charge carrier clouds is much smaller than the thickness of the detector, the origin on the $x$-axis can be chosen to be at the center of the clouds and the electrodes can be regarded as located at $\pm\infty$; (2) the external electric field $E_\text{ex}$ can be regarded as a constant in the region around the clouds and is parallel to the $x$-axis; (3) the charge carrier clouds are simplified as a horizontal cylindrical tube with a radius of $R$; (4) the number density is assumed to be a constant along its radius and a Gaussian distribution~\cite{Sosin2012a} $\text{Exp}\left\{-x^2/(2\sigma^2)\right\}$ along $x$, where $\sigma$=R/3; (5) the value of $R$ is estimated with the amount of energy deposition and $\diff E/\diff x$ of incident particles in germanium; (6) the clouds are allowed to evolve only along $x$ under the influence of the external field $E_\text{ex}$ and the induced field $E_\text{in}$ once the electron and hole clouds are separated from each other; and (7) the diffusion of the clouds in any direction is ignored. The diffusion along $x$ was original included in the simulation but turned out to be negligible and is ignored safely for the reason of simplicity. The transverse diffusion reduces the density of the clouds and were believed to have a non-negligible negative impact on the plasma time by some authors~\cite{Taroni1969a, jba}. However, their discussion was mainly about the diffusion through the side surface of a long track in parallel to the electric field in a silicon detector. In case of low energy recoils in germanium detectors, the track length is rather short according to $\diff E/\diff x$. The plasma cloud is more like a sphere\footnote{It is regarded as a cylinder with roughly equal height and radius in this paper to simplify the numerical treatment.} than a long track. Since the diffusion along $x$ was calculated to be negligible, a big difference in transverse diffusion is not expected. Of course an accurate treatment of the diffusion is more convincing than such a simple argument, however, since our intention is to offer a first order approximation, we do not take the transverse diffusion into account.

\subsection{Effect of mobility on plasma time}
\label{s:mobi}

The mobility of charge carriers is determined primarily by the scattering of charge carriers with the following components in a germanium crystal~\cite{Debye1954a}: ionized impurities, neutral impurities, lattice phonons and dislocations. Mobilities with respect to individual scattering processes can be combined according to Matthiessen's rule:
\begin{equation}
  \frac{1}{\mu_\text{tot}} = \frac{1}{\mu_\text{ion}}+\frac{1}{\mu_{\text{others}}},
  \label{e:mu1}
\end{equation}
where $\mu_\text{tot}$ represents the total mobility, $\mu_\text{ion}$ is the contribution from the ionized impurities and $\mu_{\text{others}}$ represents the contribution from scattering processes other than $\mu_\text{ion}$. In the intrinsic region of a high-purity germanium detector, the concentration of ionized impurities is so low that $\mu_\text{ion}$ can be safely ignored at 77 Kelvin~\cite{brown1962}. In this case, $\mu_\text{tot} \approx \mu_\text{others}$ and the measured value of $\mu_\text{tot}$ along $\langle100\rangle$ direction, 40180~cm$^2$/(V$\cdot$s) for electrons~\cite{mihailescu2000} and 66333~cm$^2$/(V$\cdot$s) for holes~\cite{reggiani1977}, can be used as an approximation of $\mu_\text{others}$.

In all existing studies of the plasma time in semiconductor detectors, the mobility is treated as a constant~\cite{wse,Tove1967a,Taroni1969a}. This is not necessarily the case when the charge carrier concentration is too high. Consider a \SI{1}{\kilo\eV} NR, the average track length is about \SI{2e-7}{cm} based on the stopping power model in Ref.~\cite{mei2}. Assume the plasma cloud takes the shape of a cylindrical tube with its initial radius and height equal to the average track length, the initial average charge carrier concentration is then about \SI{2e21}{/cm^3}. With such a high concentration, electrons and holes in the cloud would work as ionized impurities and slow down the drift velocity of themselves. In this case, the contribution of $\mu_\text{ion}$ cannot be ignored and has to be properly estimated.

When the ionized impurity concentration, $N$, is in the range of $[10^{14}, 10^{18}]$/cm$^3$, $\mu_\text{ion}$ can be calculated based on the Brooks-Herring (BH) model~\cite{Chat1981a}:
\begin{equation}
  \mu_\text{BH}=\frac{128\sqrt{2\pi}(\varepsilon_\text{Ge}\varepsilon_0)^{2}(k_BT)^{3/2}}{m^{\ast 1/2}NZ^2q^3}/ln\frac{24m^{\ast}\varepsilon_\text{Ge}\varepsilon_0(k_BT)^2}{Nq^2\hbar^2},
  \label{e:muBH}
\end{equation}
where $k_B$ is the Boltzmann constant, $T$ is the temperature, $m^{\ast}$ is the band-edge effective mass equal to 0.12$m_0$ for electrons and equal to 0.21$m_0$ for holes with $m_0$ the mass of electron, $Z$ is the charge of the impurity in the unit of electron charge, and $\hbar$ is the reduced Planck constant.

When $N$ is in the range of $[4\times 10 ^{18}, 8 \times 10^{20}]$/cm$^3$, the germanium crystal can be simply treated as a conductor. In this case, $\mu_\text{ion}$ is the same as the mobility in the conductor ($\mu_\text{C}$) and can be related to the resistivity through the following relationship:
\begin{equation}
  \mu_\text{C} =\frac{1}{qN\rho},
  \label{e:muC}
\end{equation}
where $\rho$ is the resistivity in Ohm$\cdot$cm. The data for $\rho$ was provided by Sze and Irvin~\cite{Sze1968a}.

\begin{figure} [htbp]
  \centering
  \includegraphics[trim={0.7cm 0.4cm 1.4cm 1.1cm},clip,width=0.49\linewidth]{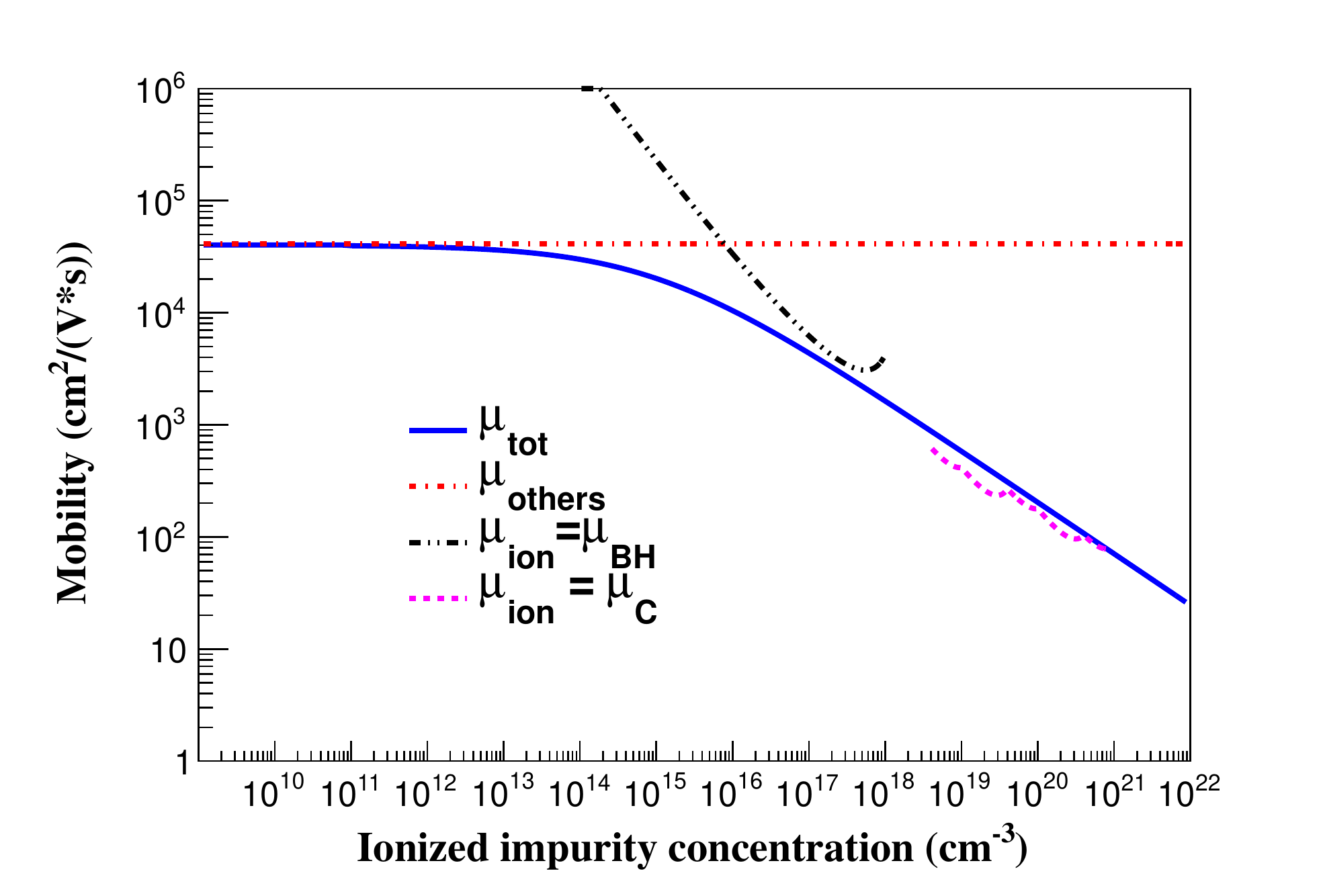}
  \includegraphics[trim={0.7cm 0.4cm 1.4cm 1.1cm},clip,width=0.49\linewidth]{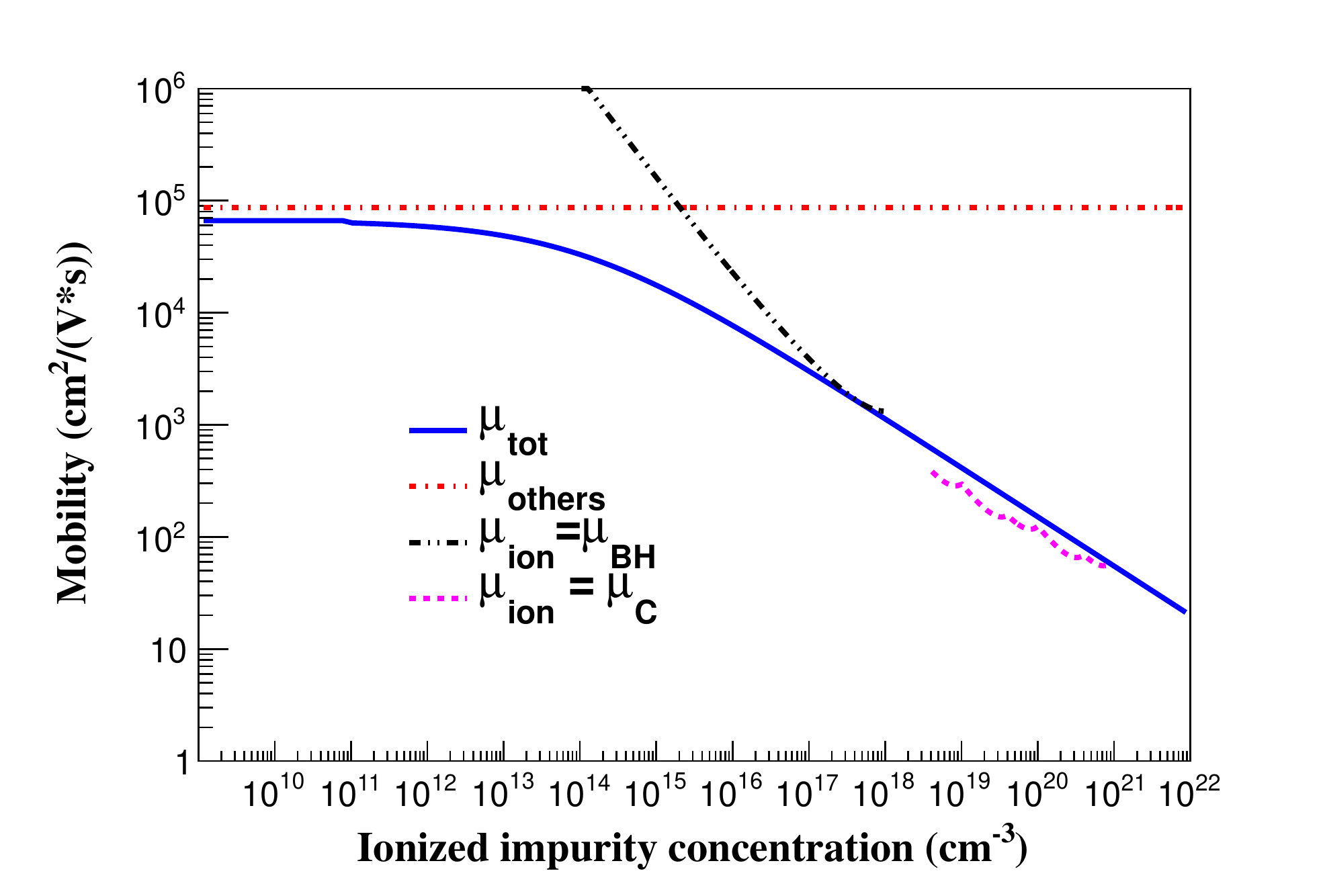}
  \caption{Electron (left) and hole (right) mobilities as a function of ionized impurity concentration.}
  \label{f:mu}
\end{figure}

There are no measurements or models available to evaluate $\mu_\text{ion}$ in the region of $[1\times 10^{18}, 4\times 10^{18}]$/cm$^3$. A natural way to get the transition curve would be to simply connect the end point of $\mu_\text{BH}$ to the start point of $\mu_\text{C}$. However, this treatment introduces sudden changes in the derivatives of $\mu_\text{ion}$, which causes artificial oscillations in numerically calculated distributions. To avoid this problem, we assume $\mu_\text{ion}$ can be described by equation~\ref{e:muE} and \ref{e:muH}. The parameters in these equations were set by hand so that the curves can represent a smooth transition from $\mu_\text{BH}$ to $\mu_\text{C}$.
\begin{equation}
\mu_\text{ion} = 10^{-0.46logN+11.51}, \;\text {\it for electrons,}
\label{e:muE}
\end{equation}
\begin{equation}
\mu_\text{ion} = 10^{-0.44logN+10.98}, \;\text {\it for holes.}
\label{e:muH}
\end{equation}
The total mobility calculated with eq.~\ref{e:mu1} as a function of the ionized impurity concentration for electrons and holes are shown in figure~\ref{f:mu}.

\subsection{Evolution of various distributions}

\begin{figure} [htbp]
  \centering
  \includegraphics[width=0.6\linewidth]{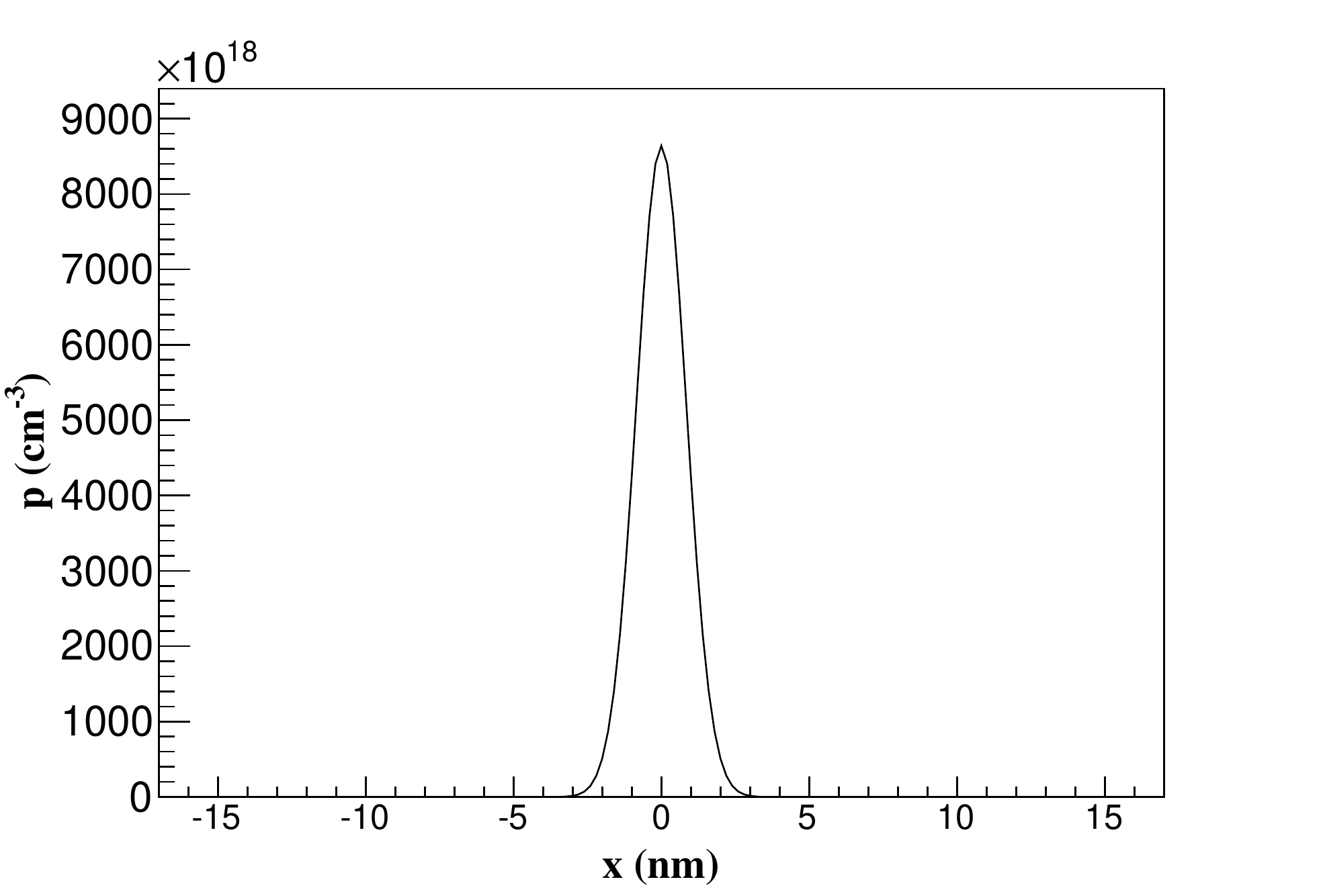}
  \caption{The number density distribution of holes created by a 5~keV NR after 0.01~ns of evolution under a field strength of 500~V/cm.}
  \label{f:pDis}
\end{figure}

\begin{figure} [htbp]
  \centering
  \includegraphics[width=0.6\linewidth]{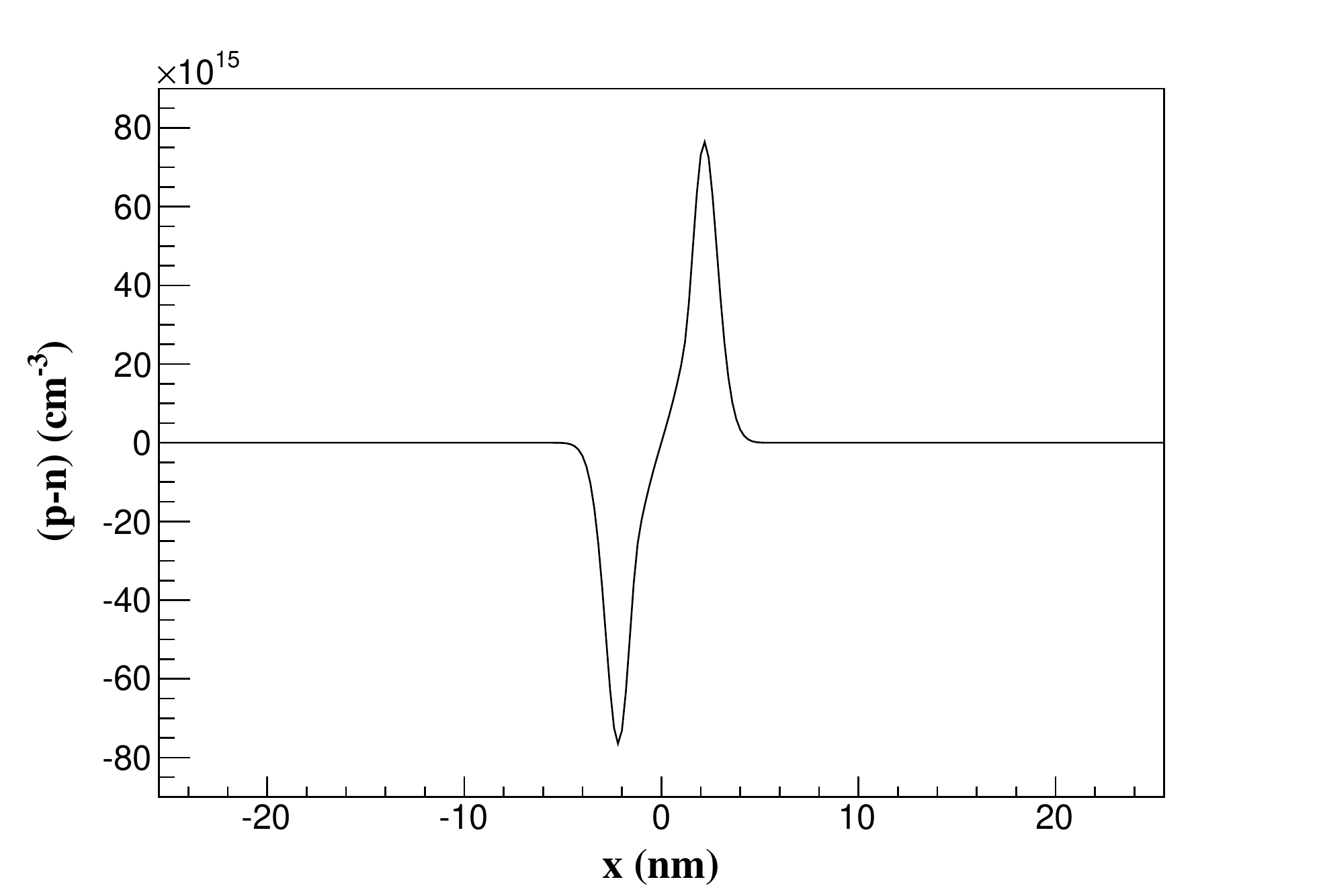}
  \caption{The difference of the number density distributions between holes and electrons, $p(\V{x})-n(\V{x})$, after \SI{1e-8}{ns} of evolution under the same condition of figure~\ref{f:pDis}.}
  \label{f:p_nDis}
\end{figure}

Figure~\ref{f:pDis} shows the number density distribution of holes created by a 5~keV NR after 0.01~ns of evolution under a field strength of 500~V/cm. It does not differ much from the initial Gaussian distribution. However, the evolution is already visible in an earlier stage (after only \SI{1e-7}{ns}) in figure~\ref{f:p_nDis}, where the difference of the number density distributions between holes and electrons, $p(\V{x})-n(\V{x})$, is shown. Tiny amount of electrons and holes are eroded out of the plasma zone on the edges of the Gaussian distribution by external electric field. Note that the evolution time is calculated based on the number of steps and the time interval of each step ($\diff t$). In our numerical calculation, $\diff t$ = \SI{1e-9}{ns} for the first 100 steps of the evolution, $\diff t$ = \SI{1e-7}{ns} for the next 1000 steps and $\diff t$ = \SI{5e-7}{ns} for the rest.

\begin{figure} [htbp]
  \centering
  \includegraphics[width=0.6\linewidth]{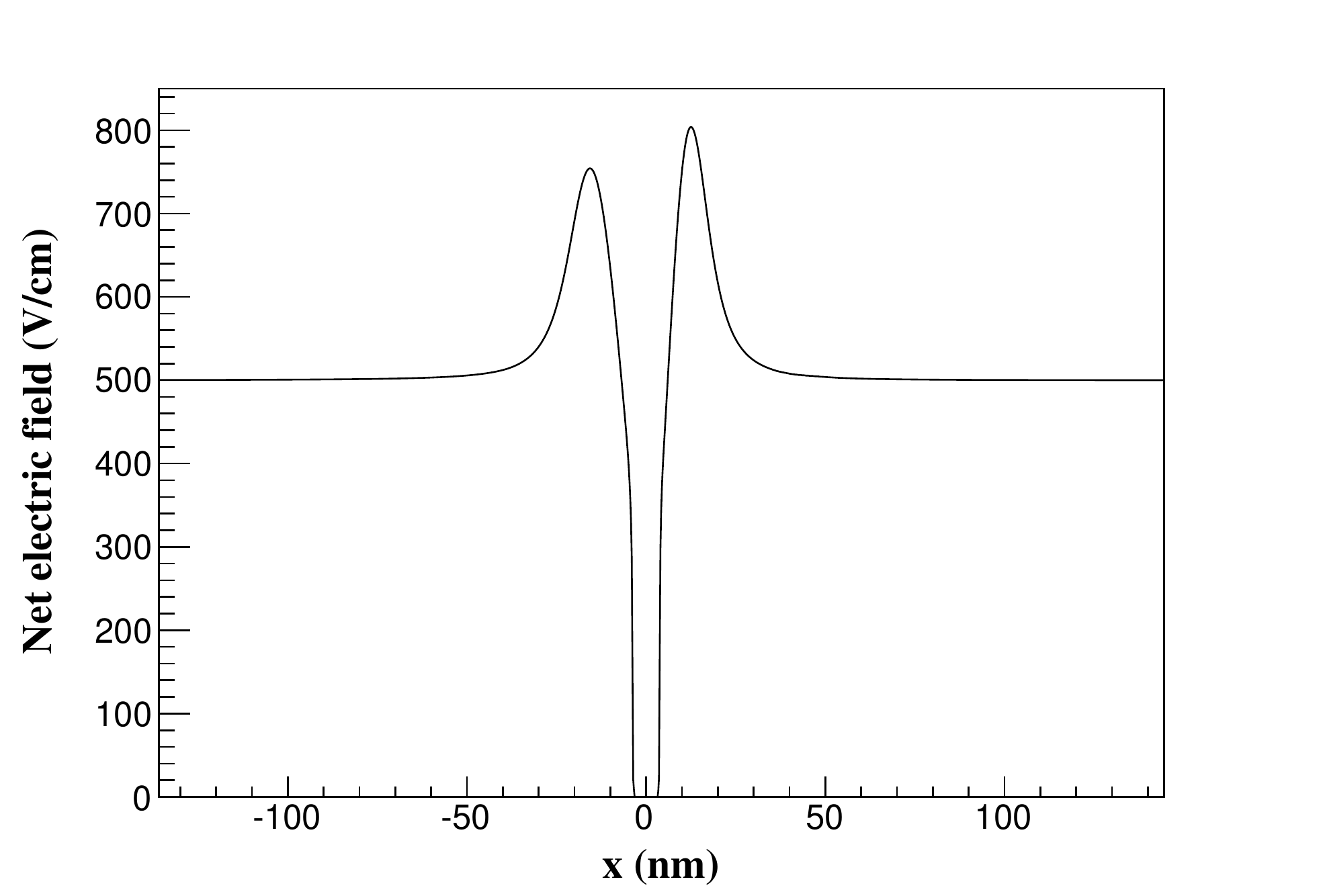}
  \caption{The net electric field distribution after \SI{1.5e-4}{ns} of evolution under the same condition of figure~\ref{f:pDis}.}
  \label{f:netEDis}
\end{figure}

Figure~\ref{f:netEDis} shows the net electric field distribution after \SI{1.5e-4}{ns} of evolution under the same condition of figure~\ref{f:pDis} and \ref{f:p_nDis}. The zero field region around the center of the plasma zone results from the screening of the external electric field by net electrons and holes accumulated on the edges of the plasma zone shown in figure~\ref{f:p_nDis}. The two peaks around the valley are due to the fact that the induced electric field outside the plasma zone is in parallel with the external electric field. Their difference in height is due to the difference of mobilities between electrons and holes.

\begin{figure} [htbp]
  \centering
  \includegraphics[width=0.6\linewidth]{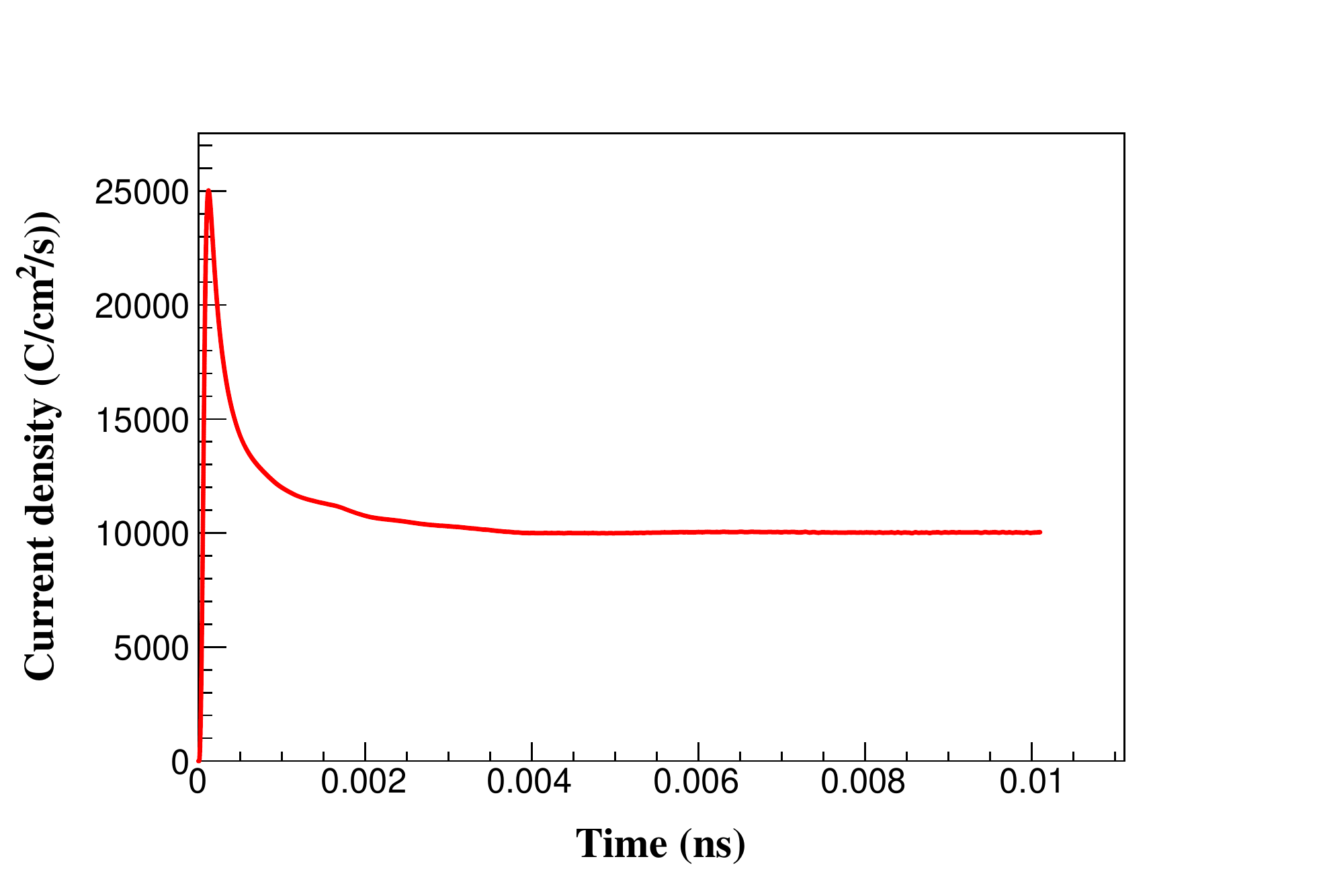}
  \caption{Evolution of overall current density \SI{10}{nm} to the right of the origin within 0.01~ns.}
  \label{f:jt}
\end{figure}

Figure~\ref{f:jt} shows the evolution of overall charge current density on the right edge of the Gaussian distribution (\SI{10}{nm} away from the center) within 0.01~ns under the same condition of figure~\ref{f:pDis}. After some initial fluctuations, it quickly approaches a constant value. This constant is the so-called \emph{steady-state erosion current density} as defined in Tove and Seibt's work~\cite{Tove1967a}. It can be understood as that charge carriers inside the plasma zone do not move much, they can only be slowly eroded away from the edges of the plasma zone, and such condition would not change until there are not enough charge carriers left in the plasma zone to support the magnitude of the current density.

\subsection{Estimation of plasma time}
Eq.~\ref{e:tp} shows ideally how the plasma time can be estimated. However, the results shown in this work are not calculated that way due to the following two reasons: First, the minimal time interval is chosen to be \SI{1e-9}{ns}, otherwise the calculation is not precise enough to represent the real evolution. It requires about \num{1e7} iterations to reveal the current density shown in figure~\ref{f:jt}. The distributions shown in the previous section do not change much in a long period once the current density approaches the constant value. Since the calculation was quite time consuming, it was stopped when the current density became effectively constant. Secondly, as shown in eq.~\ref{e:ce} and \ref{e:ch}, numerical differentiation is involved in the calculation. Small rounding errors propagate over iterations and become too large after too many iterations. The calculation had to be stopped before that.

Due to the fact that the one-dimensional current density, $j$, reaches a constant value almost immediately, the plasma time, $t_{pl}$, can then be estimated using the following relation instead:
\begin{equation}
  t_{pl} = Q/(jA),
  \label{e:pt}
\end{equation}
where, $Q$ = $qE/\varepsilon$ is the initial total charge created by a recoil event, with $E$ the electronic-equivalent recoil energy and $\varepsilon = 3$~\si{\eV} the average energy expended per e-h pair for germanium at 77 Kelvin~\cite{E&R, shock, pehl, klein1, klein2, A&B}, and $A$ is the cross-section of the electron or hole clouds shown in figure~\ref{f:cfg}, $A = \pi R^2$ with $R=\frac{E}{\diff E/\diff x}$.

\section{Results of the numerical calculation}
\label{s:res}
Table~\ref{t:nr} and \ref{t:er} summarize the calculated plasma times of NRs at \SIlist{1; 5; 10; 50}{\kilo\eV} and ERs at \SIlist{0.17; 1.08; 2.36; 15.15}{\kilo\eV} at eight different external fields, \SIlist{100; 150; 200; 250; 350; 500; 750; 1000}{V/cm}. The energies of ERs were chosen such that they were the same as the visible energies of the NRs, which were calculated using the ionization efficiency given by the Lindhard's theory~\cite{lind10}. 
\begin{table}[htbp]
  \centering
  \caption{The plasma time in ns for NRs.}
  \label{t:nr}
  \setlength{\extrarowheight}{0.1cm}
  \begin{tabular}{|c|c|c|c|c|} \hline
    \backslashbox{Field}{Energy/keV}& 1 & 5 & 10 & 50 \\ \hline 
    100~V/cm &212.0 &345.4 &416.6 &610.1\\ \hline
    150~V/cm &110.3 &189.1 &224.7 &298.9\\ \hline
    200~V/cm &69.50 &123.0 &146.1 &183.6\\ \hline
    250~V/cm &48.82 &87.18 &104.1 &138.7\\ \hline
    350~V/cm &28.71 &50.90 &63.39 &72.26\\ \hline
    500~V/cm &16.85 &28.95 &32.82 &36.86\\ \hline
    750~V/cm & 8.89 &15.56 &17.98 &20.84\\ \hline
    1000~V/cm &5.64 &10.12 &11.71 &14.12\\ \hline
  \end{tabular}
\end{table}

\begin{table}[htbp]
  \centering
  \caption{The plasma time in ns for ERs.}
  \label{t:er}
  \begin{tabular}{|c|c|c|c|c|} \hline
    \backslashbox{Field}{Energy/keV}& 0.17 & 1.08 & 2.36 & 15.15 \\ \hline 
    100~V/cm &395.3 &19.27 &10.58 &2.54\\ \hline
    150~V/cm &214.4 & 9.66 & 5.58 &1.28\\ \hline
    200~V/cm &142.9 & 6.19 & 3.18 &0.76\\ \hline
    250~V/cm &103.1 & 4.21 & 2.19 &0.51\\ \hline
    350~V/cm &62.81 & 2.45 & 1.24 &0.30\\ \hline
    500~V/cm &34.86 & 1.47 & 0.69 &0.16\\ \hline
    750~V/cm &19.27 & 0.69 & 0.35 &0.09\\ \hline
    1000~V/cm&12.61 & 0.43 & 0.22 &0.06\\ \hline
  \end{tabular}
\end{table}

\begin{figure} [htbp]
  \centering
  \includegraphics[width=0.6\linewidth]{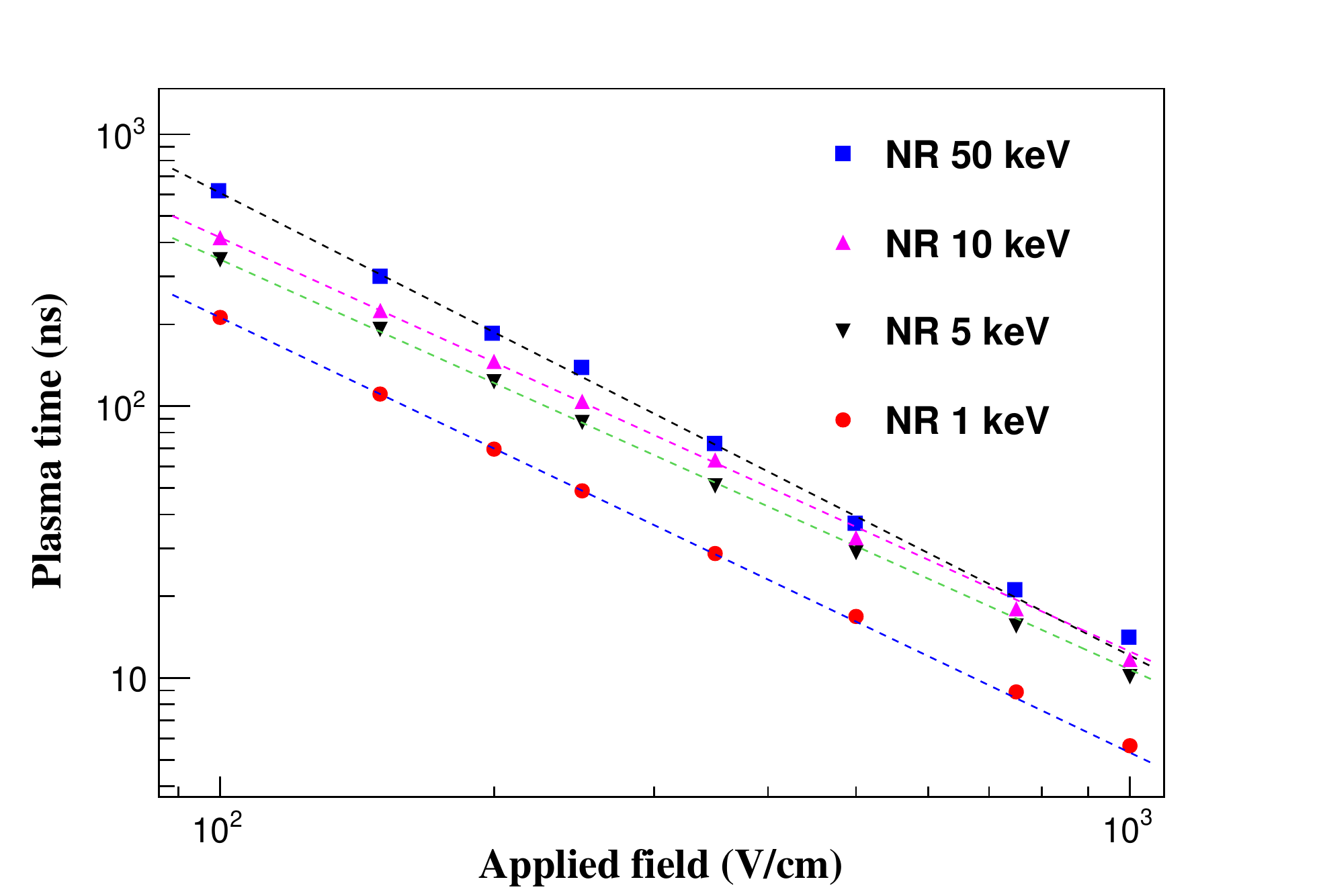}
  \caption{The plasma time versus the applied field for NRs with energies, 1 keV, 5 keV, 10 keV and 50 keV.}
  \label{f:pt_NR}
\end{figure}

\begin{figure} [htbp]
  \centering
  \includegraphics[width=0.6\linewidth]{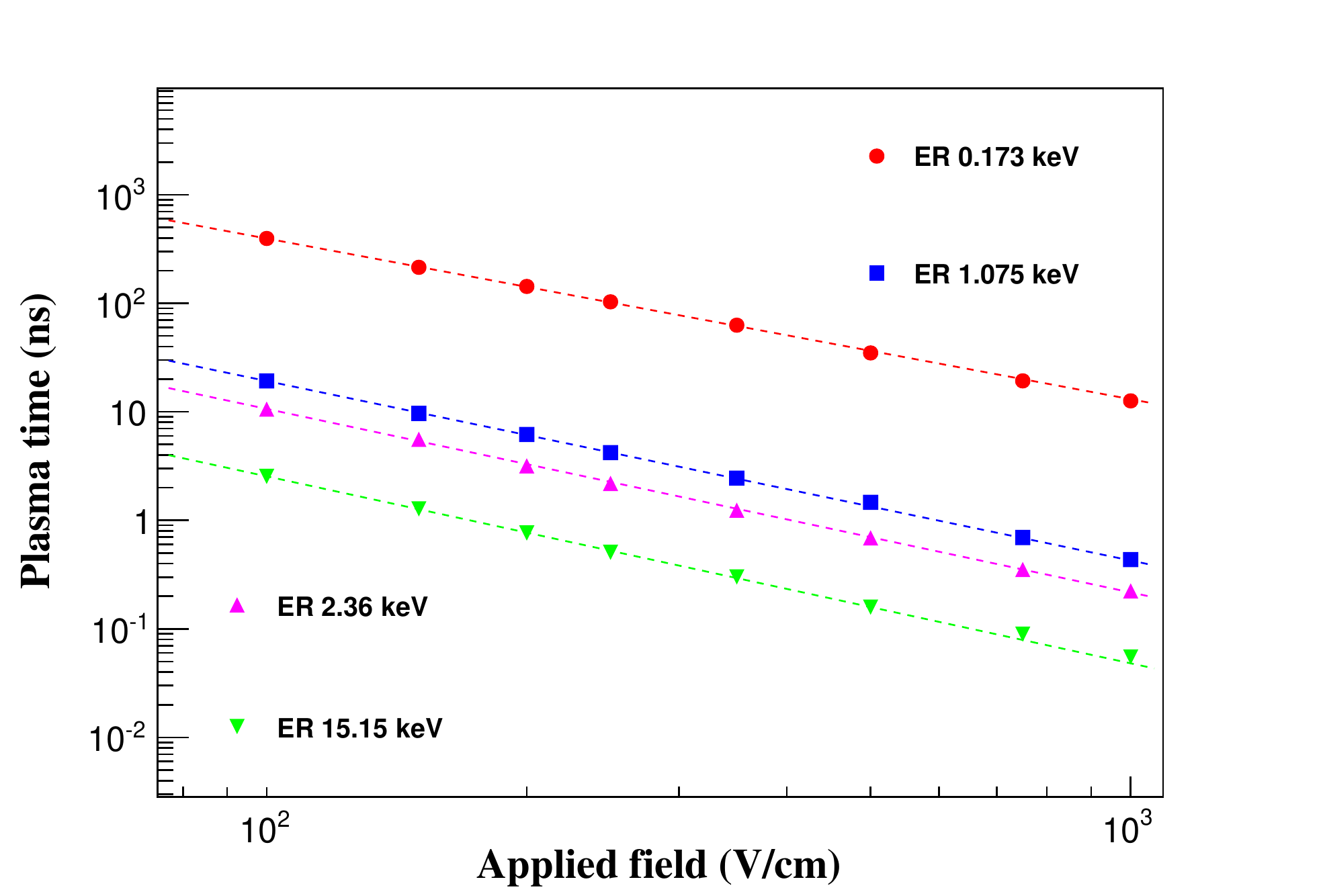}
  \caption{The plasma time versus the applied field for ERs with energies, 0.173 keV, 1.075keV, 2.36 keV and 15.15 keV.}
  \label{f:pt_ER}
\end{figure}

Figure~\ref{f:pt_NR} and \ref{f:pt_ER} show the plasma time as a function of the applied field for NRs and ERs, respectively, based on the data in table~\ref{t:nr} and \ref{t:er}. The plasma times for both NRs and ERs are inversely proportional to the applied field. However, the plasma time for NRs increases as the recoil energy increases, while the plasma time for ERs decreases as the recoil energy increases. This is because the stopping power ($\diff E/\diff x$) is the dominant factor in determining the plasma time in eq.~\ref{e:pt}, and $\diff E/\diff x$ increases as NR energy increases while it decreases as ER energy increases. The best-fit function in figure~\ref{f:pt_NR} and figure~\ref{f:pt_ER} is, $t_{pl} = p_0\cdot E_{a}^{p_1}$, where $E_{a}$ is the applied field, $p_0$ and $p_1$ are the fitting parameters. 

\begin{figure} [htbp]
  \centering
  \includegraphics[width=0.6\linewidth]{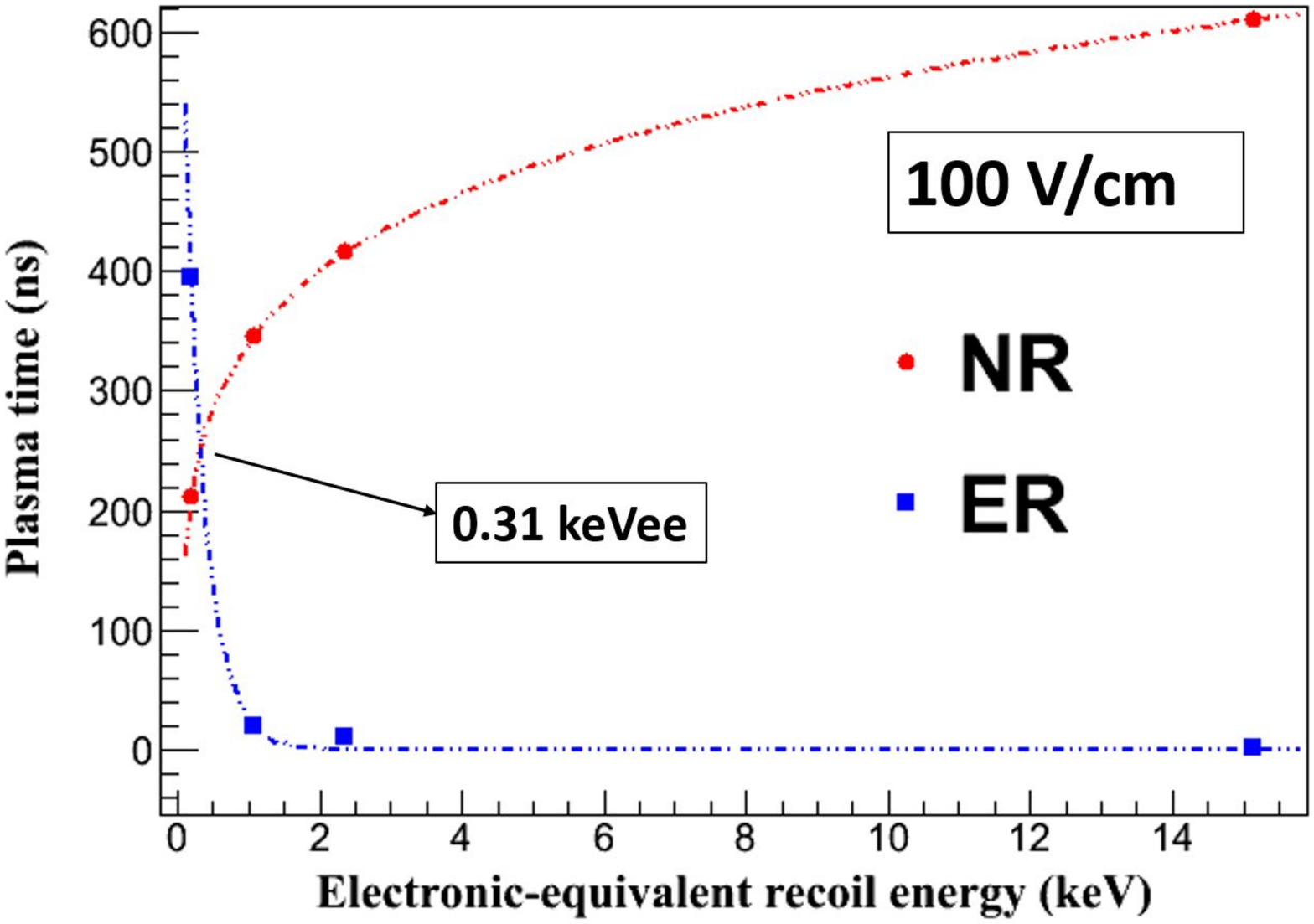}
  \caption{The discrimination of NRs from ERs with the plasma time under the applied field 100 V/cm.}
  \label{f:100}
\end{figure}

\begin{figure} [htbp]
  \centering
  \includegraphics[width=0.6\linewidth]{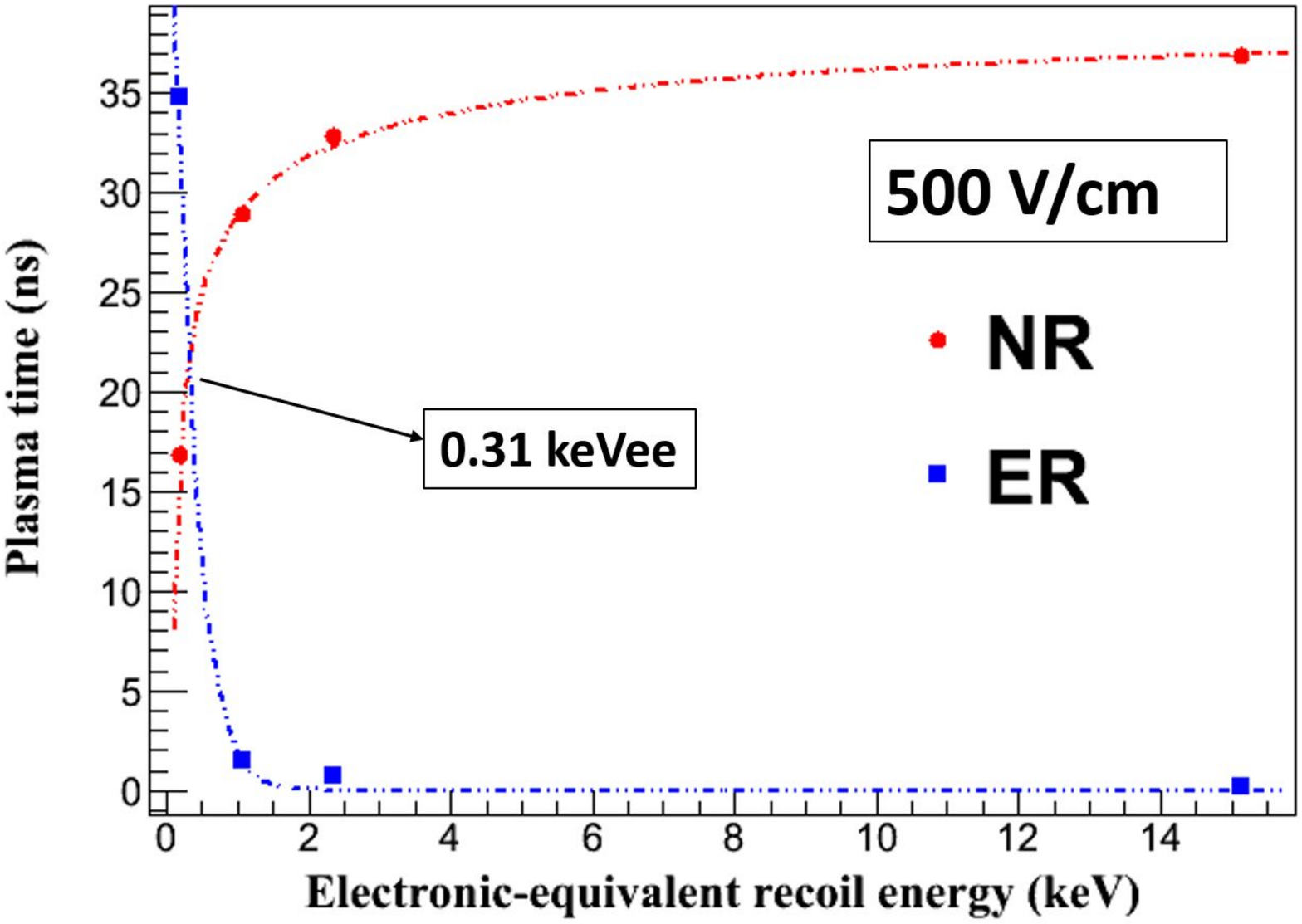}
  \caption{The discrimination of NRs from ERs with the plasma time under the applied field 500 V/cm.}
  \label{f:500}
\end{figure}

\begin{figure} [htbp]
  \centering
  \includegraphics[width=0.6\linewidth]{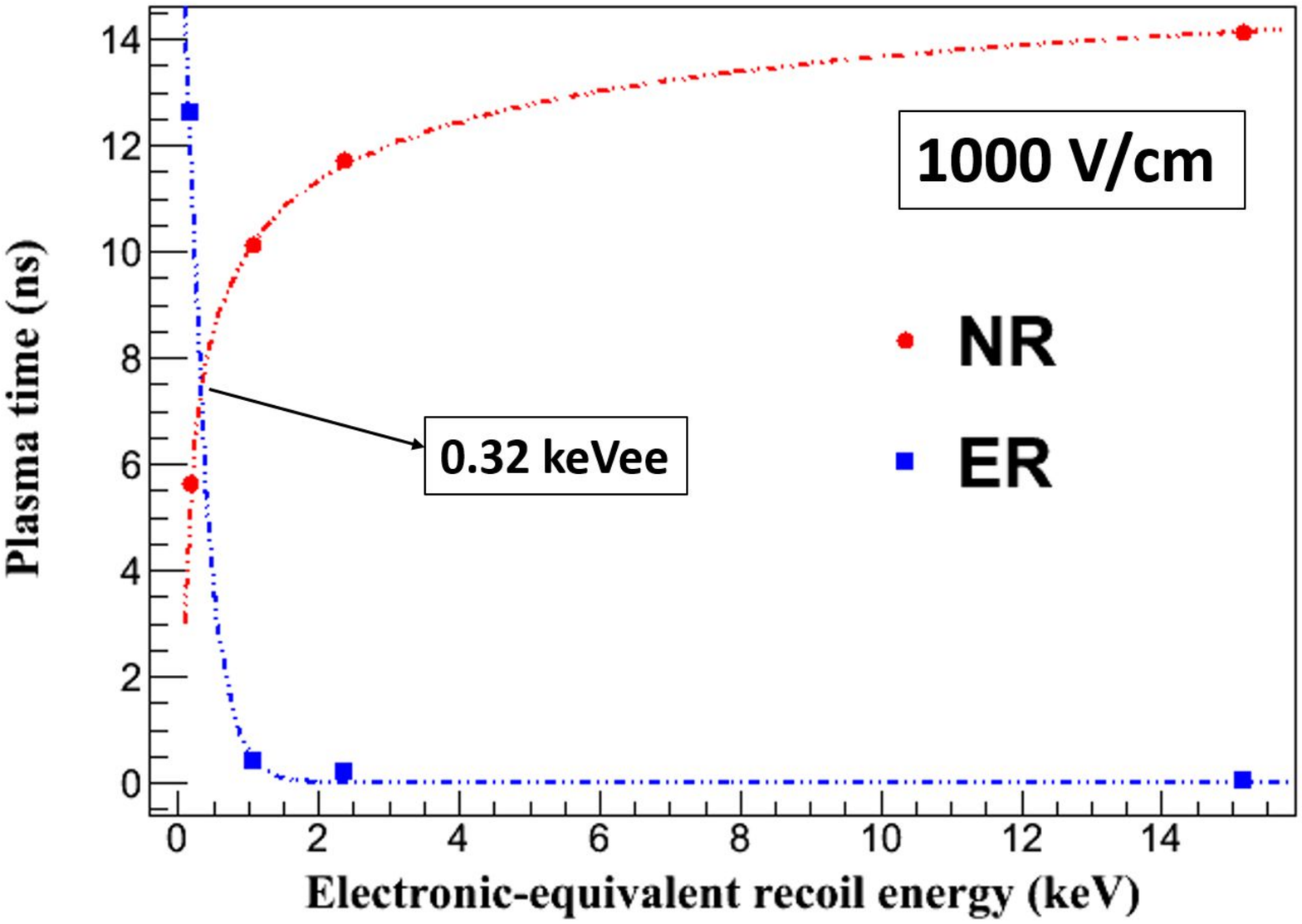}
  \caption{The discrimination of NRs from ERs with the plasma time under the applied field 1000 V/cm.}
  \label{f:1000}
\end{figure}

The capability of discriminating NRs from ERs using their differences in the plasma times in a germanium detector is investigated. Three representative applied fields, 100 V/cm, 500 V/cm and 1000 V/cm, were chosen as examples to show the discrimination power in figure~\ref{f:100}, \ref{f:500} and \ref{f:1000}, respectively.

The best-fit function for data points of NRs and ERs in figure~\ref{f:100}, figure~\ref{f:500} and figure~\ref{f:1000} are, $t_{pl}=p_0\cdot E_r^{p_1}+p_2$ (NRs) and $t_{pl} = p_3^{E_r+p_4}$ (ERs), with $E_r$ the electronic-equivalent recoil energy, $p_0$, $p_1$, $p_2$, $p_3$ and $p_4$ the fitting parameters. The values of these fitting parameters are listed in table~\ref{t:nr_pa} and table~\ref{t:er_pa} for NRs and ERs, respectively. Note that, as shown in table~\ref{t:nr_pa} and table~\ref{t:er_pa}, the value of $\chi^2/ndf$ for most of fits is quite small. This is mainly due to the fact that no error bars are introduced when fitting the data points in figure~\ref{f:100}, figure~\ref{f:500} and figure~\ref{f:1000}.
\begin{table}[htbp]
  \centering
  \caption{The fitting parameters for the fits of NRs in figure~\ref{f:100}, figure~\ref{f:500} and figure~\ref{f:1000}.}
  \label{t:nr_pa}
  \begin{tabular}{|c|c|c|c|c|} \hline
    & $p_0$ & $p_1$ & $p_2$ & $\chi^2$/ndf \\ \hline 
    figure 9 & 619.2$\pm$39.28&0.13$\pm$0.0076 &-278.6$\pm$38.64 &2.6/1 \\ \hline
    figure 10 & -12.28$\pm$2.03 &-0.39$\pm$0.057 &41.21$\pm$1.74 &0.27/1 \\ \hline
    figure 11 & -8.69$\pm$0.96 &-0.23$\pm$0.025 &18.74$\pm$0.9 &0.012/1  \\ \hline    
  \end{tabular}
\end{table}

\begin{table}[htbp]
  \centering
  \caption{The fitting parameters for the fits of ERs in figure~\ref{f:100}, figure~\ref{f:500} and figure~\ref{f:1000}.}
  \label{t:er_pa}
  \begin{tabular}{|c|c|c|c|} \hline
    & $p_3$ & $p_4$ & $\chi^2$/ndf \\ \hline 
    figure 9 & 0.036$\pm$0.01 &-1.97$\pm$0.22 &112.8/2 \\ \hline
    figure 10 & 0.03$\pm$0.01 &-1.19$\pm$0.1 &0.48/2 \\ \hline
    figure 11 & 0.024$\pm$0.009 &-0.85$\pm$0.068 &0.05/2  \\ \hline    
  \end{tabular}
\end{table}

Using the two fitting functions mentioned above for data points in figure~\ref{f:100}, figure~\ref{f:500} and figure~\ref{f:1000}, we found out that only in a region around $\sim$0.3 keVee no discrimination is possible for a generic germanium detector utilizing the plasma time. Note that the plasma effect, in general, can be observed by measuring the plasma time and the amplitude distortion of the pulse shape due to the recombination of charge carriers induced by plasma time. However, the lifetime of electrons in germanium at 77 Kelvin is above 10$^{-4}$ seconds, the recombination of charge carriers within the plasma time of less than 100 nanoseconds is negligible according to the recombination probability function developed in~\cite{wangmei}. 

\section{Experimental consideration on measuring the plasma effect}
\label{s:expe}
High-purity germanium detectors are commonly operated at a field strength of 1000~V/cm. As shown in figure~\ref{f:1000}, the difference of the plasma times between NRs and ERs is around 10~ns in this case. The charge carrier drift time is several hundred nanoseconds in the case of coaxial detectors and more than \SI{1}{\micro\second} in the case of point-contact ones. Such a long drift time washes out the subtle difference due to the plasma effect. Besides, pre-amplifiers with bandwidths around 350~MHz and digitizers with sampling rates about 1~GHz are needed to resolve time structure in nano second range. Such electronics are not commonly used in germanium detector systems. These are why the plasma effect in germanium has not yet been observed.

A successful measurement of the plasma effect in germanium detector requires a substantial decrease of the drift time and a significant increase of the plasma time.  The increase of the plasma time can be achieved by simply reducing the external field strength. However, there is a lower limit of such a reduction, that is, the field must be strong enough to deplete the detector. One way to reduce the depletion voltage of a detector is to make it thinner. A planar detector is hence a better choice than a coaxial one. Another way is to operate a detector at low enough temperatures, where most ionized impurities freeze out and there is no need to have very high voltage to swipe out space charges. The reduction of the drift time can be achieved by both reducing the drift length and increasing the charge carrier drift mobility, which increases rapidly when the temperature goes down, since the lattice scattering becomes less frequent~\cite{brown1962, otta75}. Liquid neon would be a better choice than liquid nitrogen as a cooling medium, given its lower boiling point, 27.07 Kelvin. By operating a thin planar germanium detector at liquid neon temperature, it is possible to deplete the detector at about 100~V to achieve a drift time of about 10~ns and a plasma time difference of about 400~ns. Such a big difference can be easily measured using electronics with moderate bandwidths. Pre-amplifiers with a rise time less than 10~ns have been developed for the GERDA experiment~\cite{preamp}, which makes it possible to resolve subtler differences in plasma time at higher depletion voltages or around 0.3~keVee energy region.

There are several other advantages coming from the use of liquid neon as cooling material. First of all, as other noble gas elements, liquid neon is relatively easy to purify, a key requirement in dark matter experiments. Secondly, there is no long term radioactive isotope. Third, it emits scintillation light, providing an anti-coincident veto for dark matter measurement. Last but not least, it is available in large quantities and is relatively inexpensive, which are favorable for large scale experiments.

\section{Conclusion}
\label{s:conc}
We have conducted a numerical calculation of the plasma time for both NRs and ERs down to 1 keV. The plasma time difference is in the range of a few nanoseconds to a few hundred nanoseconds depending on the recoil energy and the applied electric field for NRs and ERs. If one uses a lower applied electric field (100 V/cm), the difference in the plasma time between NRs and ERs can be enhanced. This difference in the plasma time will result in a difference in the rise time of the pulse shapes for a generic germanium detector with a good timing resolution at a level of $\sim$1 to $\sim$10 ns at 77 Kelvin. This particular time difference induced by the plasma effect can be used to discriminate NRs from ERs for a generic germanium detector with appropriate design for the geometry and electric field for the direct detection of rare physics processes.

\acknowledgments
The authors wish to thank Christina Keller for her careful reading of this manuscript.  This work was supported in part by NSF PHY-0919278, NSF PHY-1242640, NSF OIA 1434142, DOE grant DE-FG02-10ER46709, the Office of Research at the University of South Dakota and a research center supported by the State of South Dakota. Computations supporting this project were performed on High Performance Computing systems at the University of South Dakota. We thank its manager, Doug Jennewein, for providing valuable technical expertise to this project.







\begin{thebibliography}{99}
\bibitem{cdms} R. Agnese et al. (SuperCDMS Collaboration), \emph{Search for Low-Mass Weakly Interacting Massive Particles with SuperCDMS}, \emph{Phys. Rev. Lett.} {\bf 112} (2014) 241302. arXiv:1402.7137v2.
\bibitem{cog} C.E. Aalseth et al. (CoGeNT Collaboration), \emph{Results from a Search for Light-Mass Dark Matter with a P-type Point Contact Germanium Detector}, \emph{Phys. Rev. Lett.} {\bf 106} (2011) 131301.
\bibitem{schol} K. Scholberg, \emph{Coherent elastic neutrino-nucleus scattering}, \emph{Journal of Physics: Conference Series} {\bf 606} (2015) 012010.
\bibitem{bron} A. Broniatowski et al., \emph{A new high-background-rejection dark matter Ge cryogenic detector}, \emph{Phys. Lett. B} {\bf 681} (2009) 305.
\bibitem{caj} C.A.J. Ammerlaan, R.F. Rumphorst and L.A.Ch. Koerts, \emph{Particle identification by pulse shape discrimination in the p-i-n type semiconductor detector}, \emph{Nucl. Instrum. Meth.} {\bf 22} (1963) 189.
\bibitem{aaq} A. Alberigi Quaranta, M. Martini, G. Ottaviani and G. Zanarini, \emph{Proton-deuteron discrimination with a single semiconductor detector}, \emph{Nucl. Instrum. Meth.} {\bf 57} (1967) 131.
\bibitem{wde} W.-D. Emmerich, K. Frank, A. Hofmann, A. Dittner, J.W. Klein and R. Stock, \emph{Pulse-shape discrimination with surface barrier detectors}, \emph{Nucl. Instrum. Meth.} {\bf 83} (1970) 131.
\bibitem{tki} T. Kitahara, H. Geissel, S. Hofmann, G. Munzenberg and P. Armbruster, \emph{Rise-time discrimination between heavy ions and alpha particles with semiconductor detectors}, \emph{Nucl. Instrum. Meth.} {\bf 178} (1980) 201.
\bibitem{jba} J.B.A. England and G.M. Field, \emph{Z-identification of charged particles by signal risetime in silicon surface barrier detectors}, \emph{Nucl. Instrum. Meth. A} {\bf 280} (1989) 291.
\bibitem{ssk} S.S. Klein and H.A. Rijken, \emph{Pulse shape discrimination in elastic recoil detection and nuclear reaction analysis}, \emph{Nucl. Instrum. Meth. B} {\bf 66} (1992) 393.
\bibitem{gpa} G. Pausch, W. Bohne and D. Hilscher, \emph{Particle identification in solid-state detectors by means of pulse-shape analysis - results of computer simulations}, \emph{Nucl. Instrum. Meth. A} {\bf 337} (1994) 573.
\bibitem{wse} W. Seibt, K. Sundstrom and P. Tove, \emph{Charge collection in silicon detectors for strongly ionizing particles}, \emph{Nucl. Instrum. Meth.} {\bf 113} (1973) 317.
\bibitem{ecf} E.C. Finch, M. Asghar and M. Forte, \emph{Plasma and recombination effects in the fission fragment pulse height defect in a surface barrier detector}, \emph{Nucl. Instrum. Meth.} {\bf 163} (1979) 467.
\bibitem{ika} I. Kanno, \emph{Models of formation and erosion of a plasma column in a silicon surface-barrier detector}, \emph{Rev. Sci. Instrum.} {\bf 58} (1987) 1926.
\bibitem{LUX} D. S. Akerib et al. (LUX Collaboration), \emph{First Results from the LUX Dark Matter Experiment at the Sanford Underground Research Facility}, \emph{Phys. Rev. Lett.} {\bf 112} (2014) 091303.
\bibitem{xenon100} E. Aprile et al. (XENON100 Collaboration), \emph{Dark Matter Results from 225 Live Days of XENON100 Data}, \emph{Phys. Rev. Lett.} {\bf 109} (2012) 181301.
\bibitem{pandax} M. Xiao et al., \emph{First dark matter search results from the PandaX-I experiment}, \emph{Sci. China-Phys. Mech. Astron.} {\bf 57} (2014) 2024.
\bibitem{wangmei} L. Wang and D.-M. Mei, \emph{A Comprehensive Study of Low-Energy Response for Xenon-Based Dark Matter Experiments}, arXiv:1604.01083.
\bibitem{rai} R. Butsch, J. Pochodzalla and B. Heck, \emph{A direct observation of plasma delay in silicon surface barrier detectors}, \emph{Nucl. Instrum. Meth. A} {\bf 228} (1985) 586.
\bibitem{Baudis98} L. Baudis et al., \emph{High-purity germanium detector ionization pulse shapes of nuclear recoils, gamma interactions and microphonism}, \emph{Nucl. Instrum. Meth. A} {\bf 418} (1998) 348.
\bibitem{dolg} B. A. Dolgoshein et al., \emph{Electron-ion Recombination in the Track of an Ionizing Particle and the Scintillation Mechanism of Noble Gases}, \emph{Soviet Physics JETP} {\bf 29} (1969) 619.
\bibitem{lind10} J. Lindhard et al., \emph{Range Concepts
and heavy ion ranges (Notes on atomic collisions, II)}, \emph{Mat. Fys. Medd. Dan. Vid. Selsk.} {\bf 33} (14) (1963) 1.
\bibitem{Sosin2012a} Z. Sosin, \emph{Description of the plasma delay effect in silicon detectors}, arXiv:1201.2188.
\bibitem{Taroni1969a} A. Taroni and G. Zanarini, \emph{Plasma effects and charge collection time in solid state detectors}, \emph{Nucl. Instrum. Meth.} {\bf 67} (1969) 277.
\bibitem{Debye1954a} P. Debye and E. Conwell, \emph{Electrical Properties of N-Type Germanium}, \emph{Phys. Rev.} {\bf 93} (1954) 693.
\bibitem{brown1962} D. Brown and R. Bray, \emph{Analysis of Lattice and Ionized Impurity Scattering in p-Type Germanium}, \emph{Phys. Rev.} {\bf 127} (1962) 1593.
\bibitem{mihailescu2000} L. Mihailescu et al., \emph{The influence of anisotropic electron drift velocity on the signal shapes of closed-end HPGe detectors}, \emph{Nucl. Instrum. Meth. A} {\bf 447} (2000) 350.
\bibitem{reggiani1977} L. Reggiani, C. Canali, F. Nava and G. Ottaviani, \emph{Hole drift velocity in germanium}, \emph{Phys. Rev. B} {\bf 16} (1977) 2781.
\bibitem{Tove1967a} P. Tove and W. Seibt, \emph{Plasma effects in semiconductor detectors}, \emph{Nucl. Instrum. Meth.} {\bf 51} (1967) 261.
\bibitem{mei2} D.-M. Mei et al., \emph{A model of nuclear recoil scintillation efficiency in noble liquids}, \emph{Astropart. Phys.} {\bf 30} (2008) 12.
\bibitem{Chat1981a} D. Chattopadhyay and H.J. Queisser, \emph{Electron scattering by ionized impurities in semiconductors}, \emph{Rev. Mod. Phys.} {\bf 53} (1981) 745.
\bibitem{Sze1968a} S. Sze and J. Irvin, \emph{Resistivity, mobility and impurity levels in GaAs, Ge and Si at 300 K} \emph{Solid-State Electronics} {\bf 11} (1968) 599.
\bibitem{E&R} F. E. Emery and T. A. Rabson, \emph{Average energy expended per ionized electron-hole pair in silicon and germanium as a function of temperature}, \emph{Phys. Rev.} {\bf 140} (1965) A2089.
\bibitem{shock} W. Shockley, \emph{Problems related to p-n junctions in silicon}, \emph{Solid State Electronics} {\bf 2} (1961) 35.
\bibitem{pehl} R. H. Pehl et al., \emph{Accurate determination of the ionization energy in semiconductor detectors}, \emph{Nucl. Instrum. Meth.} {\bf 59} (1968) 45.
\bibitem{klein1} C. A. Klein, \emph{Semicondutor particle detectors: a research of the fano factor situation}, \emph{IEEE Transactions on Nuclear Science} {\bf 15} (1968) 214.
\bibitem{klein2} C. A. Klein, \emph{Bandgap dependence and related features of radiation ionization energies in semiconductors}, \emph{J. Appl. Phys.} {\bf 39} (1968) 2029. 
\bibitem{A&B} R. C. Alig and S. Bloom, \emph{Electron-Hole-Pair Creation Energies in Semiconductors}, \emph{Phys. Rev. Lett.} {\bf 35} (1975) 22.
\bibitem{otta75} G. Ottaviani, C. Canali and A. A. Quaranta, \emph{Charge Carrier Transport Properties of Semiconductor Materials Suitable for Nuclear Radiation Detectors}, \emph{IEEE Trans. Nucl. Sci.} {\bf 22} (1975) 192--204.
\bibitem{preamp} A. Pullia, F. Zocca, G. Pascovici, C. Boiano and R. Bassini, \emph{Ultra-fast low-noise preamplifier for bulky HPGe $\gamma$-ray sensors} \emph{IEEE Nucl. Sci. Symp. Conf. Rec.} {\bf 2005} (2005) 394--397.

\end{thebibliography}
\end{document}